\documentclass{JHEP3}
\usepackage{graphicx}
\usepackage{epsfig,amsmath,amsthm}
\newcommand{\be}{\begin{equation}}
\newcommand{\ee}{\end{equation}}
\newcommand{\bea}{\begin{eqnarray}}
\newcommand{\eea}{\end{eqnarray}}
\def\bse{\begin{subequations}}
\def\ese{\end{subequations}}
\newcommand{\IR}{\mathbb{R}} 

\def\IZ{\relax\ifmmode\hbox{Z\kern-.4em Z}\else{Z\kern-.4em Z}\fi}
\newcommand{\IS}{\mathbb{S}}

\newcommand{\non}{\nonumber \\}

\def\half{\frac{1}{2}} 

\def\del{{\partial}}

\def\bi{\begin{itemize}} \def\ei{\end{itemize}}

\def\Schw{Schwarzschild }
\def\({\left(} \def\){\right)}
\def\[{\left[} \def\]{\right]}

\title{ \center{The Delocalized Effective Degrees of Freedom \\
of a Black Hole at Low Frequencies}}

\author{Barak Kol\\
Racah Institute of Physics, Hebrew University\\
Jerusalem 91904, Israel\\
E-mail:
{\tt\href{mailto:barak_kol@phys.huji.ac.il}{barak\_kol@phys.huji.ac.il}}}

\abstract{Identifying the fundamental degrees of freedom of a
black hole poses a long-standing puzzle. Recently 
Goldberger and Rothstein forwarded  a theory of the low frequency
degrees of freedom within the effective field theory approach,
where they are relevancy-ordered but of unclear physical origin.
Here these degrees of freedom are identified with near-horizon but
non-compact gravitational perturbations which are decomposed into
delocalized multipoles. Their world-line (kinetic) action is
determined within the classical effective field theory (CLEFT)
approach and their interactions are discussed.
 The case of the long-wavelength scattering of a scalar wave off a Schwarzschild black hole
is treated in some detail, interpreting within the CLEFT approach
the equality of the leading absorption cross section with the
horizon area.}

\begin{document}

\section{Introduction}

Bekenstein's association of black-hole entropy with its horizon
area \cite{Bekenstein1973} poses a deep and long-standing
challenge to identify the fundamental (microscopic) degrees of
freedom of a black hole. In the 90's String Theory was able to
provide such a microscopical explanation of the entropy for some
black holes (near-supersymmetric) in terms of a D-brane model
\cite{SusskindUglum,StromingerVafa,CallanMaldacena} (and the
reviews \cite{revs}). Here we shall consider the low-frequency
limit of the problem which is less challenging, as it does not
require Quantum Gravity, but on the other hand we shall be able to
offer quite a full and satisfying picture.

In recent years an Effective Field Theory (EFT) approach to black
holes was introduced by Goldberger and Rothstein
\cite{GoldbergerRothstein1,Goldberger-Lect} and applied by them to
 dissipative effects \cite{GoldbergerRothstein2}. For
the purpose of describing the absorption of incoming
long-wavelength radiation by the black hole they introduce certain
black hole degrees of freedom in the spirit of effective field
theories and with no clear physical origin. They proceed to deduce
parts of the black hole effective action where these degrees of
freedom appear. Finally they speculate on their origin in terms of
a ``hypothetical two-dimensional theory localized on the stretched
horizon'' (see \cite{GR-GRF} for a further discussion). An
application of dissipative effect to spinning black holes appeared
in \cite{Porto-absorption}.

This picture raises several questions \bi
 \item What is the physical origin of the black hole horizon degrees of
 freedom, and by what procedure could they be identified?
 \item How could one obtain the effective action systematically?
 \ei

In this note we shall solve the first question, and while we make
some remarks on the second, it is mostly left for future work. We
start with explaining the main idea in section \ref{idea-section}.
We proceed in section \ref{scalar-section} to consider the
concrete case of a scalar field in the background of a \Schw black
hole, interpreting within the EFT approach the equality of the
leading absorption cross section with the horizon area, namely
$\sigma=A$. \footnote{The author is not aware of any
(asymptotically flat) black hole solution which violates this
relation.} We conclude with a short summary and discussion in
section \ref{disc-section}. Appendix \ref{app1} contains the
details of the computation of the black hole cross-section.

\section{The idea}
\label{idea-section}

Consider the full gravitational field $g_f$ in the background of a
black hole of size $r_0$. The full action is given by the usual
Einstein-Hilbert action \be
 S[g_f] = \int \sqrt{g_f}\, d^4 x\, R[g_f] ~.\ee
If we furthermore consider the scattering of long wavelength waves
off the black hole (BH), then the condition for using EFT is
fulfilled, namely that the typical length scale of the background
is much larger than the object (the BH).

Inspired by an analogy with electromagnetism
\cite{GoldbergerRothstein2} introduced the following world-line
effective action for the black hole \bea
 S_{eff}[x] &=& -m \int d\tau + \frac{\alpha}{2} \int d\tau E^{\mu\nu} E_{\mu\nu} +
  \frac{\beta}{2} \int d\tau B^{\mu\nu} B_{\mu\nu} + \dots \non
 && -\int d\tau Q^E_{ab} E^{ab} -\int d\tau Q^B_{ab} B^{ab} + \dots \label{SeffGR} \eea
where $E_{\mu\nu}:=C_{\mu\alpha\nu\beta} v^\alpha v^\beta,
~B_{\mu\nu}:=\epsilon_{\mu\alpha\beta\rho}
C^{\alpha\beta}_{~~\nu\sigma} v^\rho v^\sigma /2$ are the electric
and magnetic components of the Weyl tensor $C$; $a,b$ are frame
(``flat space'') indices rather than space-time indices; $v^\mu$
is the 4-velocity of the BH and the constants $\alpha,\beta$
appearing in the first line are yet to be determined black-hole
constants. $Q^E,\, Q^B$ are electric and magnetic quadruple type
world-line degrees of freedom, ``introducing all possible terms''
in the spirit of effective field theory even though ``the
world-line theory is not known''.

In order to learn more about the sector of the world-line theory
which involves $Q^E,\, Q^B$ \cite{GoldbergerRothstein2} analyze
the two-point function. They define a function $F(\omega)$ such
that \be
 \int dt\, e^{-i \omega t}\, \left< T Q^E(0) Q^E(t)
\right> \sim i\, F(\omega) ~.\ee
 From known results about the long
wavelength absorption cross section of \Schw they deduce \be
 {\rm Im} F(\omega) = \frac{16}{45} G^5 m^6 |\omega| ~. \ee
 Then they proceed to apply these results to compute the leading
absorption into a black hole moving through a curved background
and an inspiraling binary.

In this paper we wish to derive the horizon degrees of freedom and
their action, rather than introduce ``all possible interactions''.
The main idea is a consequence of a closer scrutiny of the concept
of ``integrating out'' within the set-up of Classical Effective
Field Theory (CLEFT) \cite{CLEFT-caged}(see \cite{NRG} for an
application of CLEFT ideas to economize through a field
re-definition the Post-Newtonian approximation).

In Quantum Field Theory we are normally cavalier about which
fields can be integrated out, and we do not make any special
requirements (except that we normally integrate out higher
frequencies before lower ones). However, the classical set-up
reveals a clear requirement. Assuming the full field decomposes as
$g_f=g_S + \bar{g}$ where $g_S$ is short wavelength components and
$\bar{g}$ is the boundary, or long-wavelength component,
\cite{CLEFT-caged} defines \be
 S_{eff}[\bar{g}] \equiv I[S,g_S] := S[\bar{g},g_S(\bar{g})] \label{Seff}\ee
namely the effective action, which is the same as integrating
$g_S$ out of $S$, is given by the full action evaluated on the
full solution which is determined by the boundary conditions
$\bar{g}$ and the equations of motion (this definition is
consistent of course with all other definitions and avoids the
Feynman path integral). This definition stresses that \emph{one is
allowed to integrate out only when the remaining fields can
specify a solution}.

In the current set-up, the boundary zones do not consist only of
the asymptotic infinity, and we must remember also the second
boundary at the horizon. Accordingly, we decompose the metric
field into three components \be
 g_f= g_H + g_S + \bar{g} \ee
 where $g_S$ is the short component of the metric field of order
$r_0$, while $g_H$ and $\bar{g}$ are longer wavelength, boundary
components: $\bar{g}$ is the usual component at asymptotic
infinity, while $g_H$ represents the component near the second
boundary, namely the horizon. Accordingly, \emph{the horizon
degrees of freedom, denoted by $Q$ in (\ref{SeffGR}) are nothing
but the horizon fields, where the $SO(3)$ representation
(indicated by the indices $a,b$) is naturally determined by the
decomposition into spherical harmonics.}

Note that so far $g_H$ was not considered within the EFT approach,
and that is justified as long as interactions between $\bar{g}$
and $g_H$ did not contribute, for instance in static situations
where we impose horizon regularity rather than allow for horizon
degrees of freedom.

\section{Free scalar field in a \Schw background}
\label{scalar-section}

We now proceed to present concrete formulae for our basic idea in
the case of a free scalar field in the \Schw background, and to
apply it to gain insight into a specific physical problem.

The action is \bea
 S &=& \int dV\, \half \( \del \phi \) ^2 \non
   &=& 2 \pi \sum_{lm} \int r^2 dr\, dt  \( f^{-1}\, \dot{\phi}^2 -f\, \phi'^{~2} -\frac{l(l+1)}{r^2}\, \phi^2 \)
   \label{S0} \eea
where in the second line we used the standard \Schw metric with
$f(r):=1-r_0/r$ and the \Schw radius $r_0$; we denoted
$\dot{\phi}:=\del_t \phi$ and $\phi'=\del_r \phi$; and moreover we
decomposed the field into spherical harmonics $\phi =\sum_{lm}
\phi_{lm} Y^{lm}$ and all quadratic expressions in $\phi$ are in a
condensed notation $\phi^2 \to \phi_{l\,m}\, \phi_{l\,-m}$.

An interesting and somewhat simpler problem is that of a 1d long
wavelength scattering off some potential $V=V(x)$ (in flat space),
but we shall not dwell on it here.

Let us consider the two boundaries: the asymptotic boundary where
$r \to \infty$ and the near horizon boundary where $r \to r_0$. At
the asymptotic boundary $f \to 1$ and we obtain \be
 S_\infty = 2 \pi \sum_{lm} \int r^2 dr\, dt  \( \dot{\phi}^2 - \phi'^{~2} -\frac{l(l+1)}{r^2}\, \phi^2 \) \ee
which is nothing but the action for a free scalar field in $\IR^4$
-- a flat 4d space-time.

At the near horizon boundary it is useful to change from the $r$
variable to the tortoise coordinate $r^*$ defined (up to a
constant) through $dr^*:=dr/f(r)$. We obtain \be
 S_H = 2 \pi \sum_{lm} \int r_0^2 dr^*\, dt  \( \dot{\phi}^2 - (\del_* \phi)^2  \)
\ee where we denoted $\del_* := \del_{r*}$ and the angular term
proportional to $f l(l+1)$ vanishes since $ f \to 0$ in this
limit. This expression describes a free scalar field on
$\IR^2_{r*,t} \times \IS^2_{r0}$ -- a flat $r^*,t$ half-plane
($r^*<0$) times a horizon sphere of radius $r_0$, but
\emph{without any kinetic term in the sphere directions}.

The essential information in the effective action is encoded by
the interactions between the fields on the boundaries of the two
asymptotic zones, the near horizon field $\phi_H$ at $r^*=0$ and
the asymptotic field $\phi_\infty$ at $r=0$. These interactions
are equivalent to prescribing gluing boundary conditions between
the two asymptotic zones and must be of the form \be
 S_{int} = 2 \pi \sum_{l} c_l\, r_0^{l+1} \int dt\, \left. \phi_H^{l}\right|_{r^*=0}\, \left. \del^l
 \phi^*_\infty\right|_{r=0}~. \label{Sint}
\ee
 This action is quadratic in the fields since the original action
(\ref{S0}) is quadratic, and thus all equations are linear. The
allowed interactions are determined by $SO(3)$ rotation
invariance, where only the traceless part of $\del^l \phi_\infty$
interacts with the spherical harmonics (the traces would have
produced redundant terms anyway, since they must vanish due to the
equations of motion). The power in the $r_0$ pre-factor is
determined by dimensional analysis and $c_l$ are dimensionless
constants to be determined through matching with a full
computation in the \Schw background.

An important feature of (\ref{Sint}) is that the terms are
relevancy-ordered in a concrete way, namely the length dimension
of each term grows with the multipole number of $\phi_\infty$.
Accordingly, the most relevant interaction comes from the monopole
of the background field $\phi_\infty$, the next correction comes
from the dipole, and so on.

\subsubsection*{Low frequency cross section equals area}

The cross section for a minimally coupled scalar field (\ref{S0})
off a 4d black hole, both rotating and non-rotating, was computed
in \cite{Page1976,Star-Chur}. Making direct use of the properties
of the confluent hypergeometric function it was found there that
in the low frequency limit the cross section equals the horizon
area \be
 \sigma = A ~. \label{sigmaA}
\ee  This result is not obvious, and somewhat surprising since
usually we think of the 2d projection of an object as a measure of
its cross section, and not its full area embedded in 3d space. As
far as the author is aware no black hole of any type provides a
counter-example.

Given the generality of the result one expects it to be a
consequence of a more general argument, rather than a specific
computation. We now proceed to show how such insight is provided
by matched asymptotic expansion (MAE) or equivalently
\cite{CLEFT-caged} CLEFT which are the appropriate tools for this
limit.

For the purpose of a MAE analysis one defines three zones:
asymptotic $r,r^* \gg 1$, intermediate $|\omega r^*| \ll 1$ and
near horizon $r^* \ll -1$, together with two overlap regions.
Limiting our attention to the leading monopole sector, and
specifying boundary conditions such that no $\phi$ radiation is
outgoing from the black hole's horizon one finds that the
absorption amplitude (or transmission through the hole's
gravitational potential) is given by \be
 T = 2 i \, \omega r_0 \ee
 (see appendix \ref{app1} for details of the calculation).

A closer scrutiny of the MAE calculation reveals that it could be
encoded \`{a} la CLEFT by a simple boundary condition requiring
that $\phi_\infty$ is regular at $r=0$ (namely the term $1/r$ is
absent, to allow continuity of
 $\phi$) \be
 \left. \phi_\infty \right|_{r=0} = {\cal O}(1)
 ~. \label{bc} \ee
This should be interpreted as the leading interaction, or boundary
condition, in the effective action (\ref{Sint}).

Altogether \emph{we obtain an insight that the universal result
(\ref{sigmaA}) is a direct consequence of the effective boundary
condition (\ref{bc})}.

\section{Discussion}
\label{disc-section}

The generalization to theories with different matter content is
now conceptually clear. Of particular interest is the case where
the space-time metric is the only field, and accordingly \emph{the
black hole degrees of freedom at low frequency consist of
near-horizon gravitational perturbations relevancy-ordered by
their multipole number}. For the \Schw space-time these
perturbations are described by the Zerilli and Regge-Wheeler
master equations, considered in the $r^* \to -\infty$ limit. The
lowest multipole in the expansion is the quadrupole, of course,
rather than the monopole as for the scalar field.

We would like to stress the general features of the emerging
picture. At low frequencies the black hole degrees of freedom were
identified with near horizon fields, namely fields on
$\IR^2_{r*,t} \times \IS^2_{r0}$, where the first factor is half
($r^*<0$) a flat 2d  space-time, and the second factor is the
horizon sphere of radius $r_0$. \emph{This space is non-compact}
(even though it does not include the asymptotic infinity) due to
the usual reason that it takes infinite asymptotic time $t$ to
reach the horizon. \emph{The origin of the degrees of freedom is
simply the available fields}, and in particular, since the black
hole is made of curved space-time its inherent degrees of freedom
are near-horizon gravitational waves. The degrees of freedom are
relevancy-ordered according to their multipole number and in this
sense they are \emph{delocalized over the horizon}. One wonders:
which of these features, if any, carries over to the fundamental,
short-distance degrees of freedom?

\subsection*{Acknowledgements}

I would like to thank the members of our group for comments on a
talk. This research is supported by The Israel Science Foundation
grant no 607/05, DIP grant H.52, EU grant MRTN-CT-2004-512194 and
the Einstein Center at the Hebrew University.

\appendix

\section{Matched asymptotic expansion for long wavelength scalar cross-section}
\label{app1}

We wish to compute the absorption (or transmission) amplitude for
an $l=0$ low frequency scalar wave impinging on a \Schw black
hole.

The full equation derived from (\ref{S0}) is \be
 -\frac{1}{r^2}\, \del_r\, f(r) r^2\, \del_r \phi = \omega^2\,
 \phi \label{fulleq} \ee
where $f=1-r_0/r$, we work in the limit $\omega\, r_0 \ll 1$ and
henceforth we use units where $r_0=1$. The boundary condition is
that there are no waves outgoing from the horizon, namely \be
\begin{array}{cclc}
 \phi &\sim& T\, \exp (i \omega r^*) & ~~~\mbox{  for  } r^* \to -\infty
 \non
 \phi &\sim& 1 \cdot \exp (i \omega r)/r + R \, \exp (-i \omega r)/r & ~~~\mbox{  for  } r \to
 +\infty \label{bc-ingoing} \end{array} \ee
where $R,T$ are unknown reflection and transmission coefficients,
which depend on $\omega$, and our objective is to determine $T$
for small $\omega$.

Our limit is appropriate for the method of matched asymptotic
expansion, since the typical length of the background field,
$1/\omega$, is much larger than $r_0$ the size of the black hole.
In order to solve this equation we divide the $r^*$ axis into
three zones whose overlap grows in the limit \bi
 \item Near horizon zone $r^* \ll -1$,

where we approximate $r \simeq 1$ and the equation (\ref{fulleq})
becomes $- \del_*^2 \phi = \omega^2\, \phi$ whose solutions are
$\exp(\pm i \omega r^*)$.

 \item Intermediate zone $\left| \omega r^* \right| \ll 1$,

where we approximate $\omega \simeq 0$ and the equation
(\ref{fulleq}) becomes $-\frac{1}{r^2}\, \del_r\, f(r) r^2\,
\del_r \phi = 0$ whose solutions are $1,\, \log(f)$.

 \item Asymptotic zone $r \simeq r^* \gg 1$,

where we approximate $f \simeq 1$ and the equation (\ref{fulleq})
becomes $-\frac{1}{r^2}\, \del_r\, r^2\, \del_r \phi = \omega^2\,
\phi$ whose solutions are $\exp(\pm i \omega r)/r$.
 \ei

{\bf Solution}. The boundary condition (\ref{bc-ingoing}) is
imposed after substituting $\phi \to T\, \phi$ for later
convenience such that in the near horizon zone we have \be
 \phi \simeq \exp (i \omega r^*) ~.\ee
We proceed to match over the overlap of the near horizon and
intermediate zones, namely for $-1/\omega \ll r^* \ll -1$ where
the equation (\ref{fulleq}) becomes $- \del_*^2 \phi =0$ and the
solutions are $1,\, r^*$ to obtain the solution in the
intermediate zone \be
 \phi \simeq 1 + i \omega \( 1 + \log{f} \) ~.\ee

Next we match over the overlap of the intermediate and asymptotic
zones, namely for $1 \ll r \ll 1/\omega$ where the equation
(\ref{fulleq}) becomes $-\frac{1}{r^2}\, \del_r\, r^2\, \del_r
\phi  =0$ and the solutions are $1,\, 1/r$ to obtain the solution
in the asymptotic zone \be
 \phi \simeq \half \(\frac{1}{i \omega} + 1 -i \omega \) \exp( i \omega r)/r
 - \half \(\frac{1}{i \omega} + 1 -i \omega \) \exp(-i \omega r)/r ~.\ee
By comparing with the asymptotic form $\phi \sim (1/T) \, \exp (i
\omega r)/r + (R/T) \, \exp (-i \omega r)/r$ we obtain the
scattering amplitudes, and in particular, the absorption (or
transmission) amplitude to leading order in $\omega$ is \be
 T = 2 i\, \omega\, r_0 ~. \ee

\end{document}